\documentclass[11pt]{article}
\usepackage{amssymb,amsmath,fullpage,graphicx}

\def\[{\begin{equation}}
\def\]{\end{equation}}

\def\sech{\mathop{\rm sech}\nolimits}

\let\^=\hat
\let\~=\tilde
\def\e{{\rm e}}
\def\d{{\rm d}}

\begin{document}
\title{From nonlocal gap solitary waves to bound states in periodic media}
\author{T.~R.~Akylas$^1$\footnote{Corresponding author, email:
trakylas@mit.edu.}, Guenbo Hwang$^2$ and Jianke Yang$^2$ \\
$^1$ Department of Mechanical Engineering, MIT, Cambridge, MA 02139\\
$^2$Department of Mathematics and Statistics, University of Vermont,
Burlington, VT 05401 }

\date{ }
\maketitle

\vspace{-0.5cm}
\begin{abstract}
Solitary waves in one-dimensional periodic media are discussed
employing the nonlinear Schr\"odinger equation with a spatially
periodic potential as a model. This equation admits two families of
gap solitons that bifurcate from the edges of Bloch bands in the
linear wave spectrum. These fundamental solitons may be positioned
only at specific locations relative to the potential; otherwise,
they become nonlocal owing to the presence of growing tails of
exponentially-small amplitude with respect to the wave peak
amplitude. Here, by matching the tails of such nonlocal solitary
waves, higher-order locally confined gap solitons, or bound states,
are constructed. Details are worked out for bound states comprising
two nonlocal solitary waves in the presence of a sinusoidal
potential. A countable set of bound-state families, characterized by
the separation distance of the two solitary waves, is found, and
each family features three distinct solution branches that bifurcate
near Bloch-band edges at small, but finite, amplitude. Power curves
associated with these solution branches are computed asymptotically
for large solitary-wave separation, and the theoretical predictions
are consistent with numerical results.
\end{abstract}

\section{Introduction}

Nonlinear wave phenomena in periodic media are currently attracting
a great deal of research interest in nonlinear optics, Bose--Einstein
condensation and applied mathematics (see Christodoulides {\it et
al.} 2003; Kivshar \& Agrawal 2003; Morsch \& Oberthaler 2006;
Skorobogatiy \& Yang 2009; Yang 2010 for reviews). Apart from
scientific curiosity, this research activity is also driven by
various potential applications, ranging from light routing in
lattice networks (Christodoulides \& Eugenieva 2001) to image
transmission through nonlinear media (Yang {\it et al.} 2011).

A characteristic feature of periodic media is the existence of bands
in the linear spectrum where linear disturbances, the so-called
Bloch modes, may propagate. Between these Bloch bands are bandgaps
in which linear disturbances are evanescent but nonlinear localized
modes, commonly known as gap solitons, are possible. Since the first
theoretical prediction (Christodoulides \& Joseph 1988) and
experimental observation (Eisenberg {\it et al.} 1998) of
fundamental solitons in the semi-infinite bandgap of one-dimensional
(1D) periodic waveguides, various types of gap solitons in one- and
multi-dimensions have been reported theoretically and
experimentally. Examples include 2D fundamental gap solitons, vortex
solitons, dipole solitons, reduced-symmetry solitons, vortex-array
solitons, truncated-Bloch-wave solitons and arbitrary-shape gap
solitons (see Yang 2010 for a review).
Multi-soliton bound states in periodic media have also been
constructed in the framework of a discrete nonlinear Schr\"odinger (NLS)
model (Kevrekidis {\it et al.} 2001). In addition, the stability of some of these gap
solitons has been examined (Pelinovsky {\it et al.} 2004,
2005; Shi {\it et al.} 2008; Yang 2010; Hwang {\it et al.} 2011).

The plethora of gap solitons in periodic media calls for
systematic identification and classification of the various types of
solutions. On this issue, some progress has
been made, particularly for gap solitons that bifurcate from linear
Bloch modes at the  edges of Bloch bands. Specifically, in 1D, only
two gap-soliton families bifurcate from each edge of a Bloch band
under self-focusing or self-defocusing nonlinearity (Neshev {\it et
al.} 2003; Pelinovsky {\it et al.} 2004; Hwang {\it et al.} 2011),
and the positions of these solitons relative to the periodic medium
(or potential) are determined by a certain recurrence relation
(Hwang {\it et al.} 2011). In 2D, four gap-soliton families (or a
multiple of four families) bifurcate from each edge of a Bloch band
under self-focusing or self-defocusing nonlinearity (Yang 2010).

However, for the majority of gap solitons, that do not bifurcate
from band edges, systematic theoretical treatment is lacking at
present.  Numerical results indicate that, typically, those solution
families feature multiple branches which bifurcate near band edges
at small (but finite) amplitude (see Yang 2010 and references
therein), yet there has been no analytical explanation for this
phenomenon.


In this article, we employ the NLS equation with a spatially
periodic potential to analytically construct and classify a broad
class of gap solitons that bifurcate away from band edges in 1D
periodic media. Our approach is based on the observation that the
two families of fundamental gap solitons that bifurcate from a band
edge can be located only at specific positions relative to the
underlying potential; otherwise, they would be nonlocal owing to the
presence of growing tails of exponentially-small amplitude with
respect to the peak wave amplitude. However, by piecing together
such nonlocal solitary waves, it is possible to construct
higher-order gap solitons, or bound states. The proposed asymptotic
theory is illustrated by working out details for bound states
involving two nonlocal solitary waves. We show that a countable set
of bound-state families can be constructed. Each family is
characterized by the separation distance of the two solitary waves
involved, and features three distinct solution branches that
bifurcate near band edges. The analytical predictions are verified
by comparison with numerical results.

\section{Preliminaries}
Our study of nonlinear wave phenomena in periodic media is based on
the 1D NLS equation,
\[
i\Psi_t+\Psi_{xx}-V(x)\Psi+\sigma \Psi^2\Psi^*=0   \label{e:NLS}
\]
with a periodic potential $V(x)$ and self-focusing ($\sigma=1$) or
self-defocusing ($\sigma=-1$) cubic nonlinearity. This equation is
the appropriate mathematical model for Bose--Einstein condensates
loaded in optical lattices (Dalfovo {\it et al.} 1999; Morsch \&
Oberthaler 2006) and laser beam transmission in photonic lattices
under the paraxial approximation (Yang 2010). Although it is
possible to consider a general periodic potential $V(x)$ as in Hwang
{\it et al.} (2011), here, for simplicity, we shall work with the
sinusoidal potential
\[
V(x)=V_0 \sin^2\hspace{-0.1cm}x\,,
\label{e:potential}
\]
which is $\pi$-periodic and also symmetric in $x$, $V_0$ being the
potential depth. This potential frequently arises in nonlinear
optics and Bose--Einstein condensates.

Solitary-wave solutions of \eqref{e:NLS} are sought in the form
\begin{equation} \label{soliton_form}
\Psi(x,t)=\psi(x)\e^{-i\mu t}\,,
\end{equation}
where $\mu$ is the propagation constant, and the amplitude function
$\psi(x)$ is real-valued and localized in space. Inserting
(\ref{soliton_form}) into \eqref{e:NLS}, we find that $\psi(x)$
satisfies
\[
\psi_{xx}-V(x)\psi+\mu \psi +\sigma \psi^3=0\,.\label{e:NLSamp}
\]
For infinitesimal solutions $\psi(x)$, the nonlinear term in the
above equation drops out. The resulting linear version of
\eqref{e:NLSamp} is a Mathieu-type equation and, by the
Bloch--Floquet theory, its wave spectrum features a band--gap
structure. Specifically, the linear version of \eqref{e:NLSamp}
admits two linearly independent solutions in the form
\[
p(x;\mu)=\e^{ikx}\~p(x;\mu)\,,\label{e:Bloch}
\]
with $\~p(x;\mu)$ being $\pi$-periodic in $x$. The wave character of
these so-called Bloch modes hinges on whether the corresponding
wavenumber $k$ is real or complex. By requiring $k$ to be real, one
then obtains an infinite number of Bloch bands for $\mu$ in which
the Bloch modes \eqref{e:Bloch} propagate. These propagation bands
are separated by gaps, where $k$ turns out to be complex, implying
evanescent behaviour.

At the edges of Bloch bands, where the modes \eqref{e:Bloch} switch
from propagating to evanescent, two Bloch modes in the band collide
at either $k=0$ or $k=\pm 1$, and a single real-valued Bloch mode
that is either $\pi$- or $2\pi$-periodic, arises there. This
suggests that edges of Bloch bands are possible bifurcation points
of solitary-wave solutions of the nonlinear equation
\eqref{e:NLSamp}. These solitary waves reside inside band gaps and
are the so-called gap solitary waves (or gap solitons).

The bifurcation of 1D gap solitons from band edges was discussed by
Pelinovsky {\it et al.} (2004) using a multiple-scale expansion in
powers of the wave amplitude, along with a certain constraint obeyed
by locally confined solutions of \eqref{e:NLSamp}. They identified
two families of gap solitons that bifurcate from each band edge,
namely `on-site' and `off-site' solitons, depending on whether the
soliton is centred at a minimum or maximum of the sinusoidal
potential \eqref{e:potential}.

More recently, Hwang {\it et al.} (2011) re-visited the problem of
small-amplitude gap solitons near band edges, taking a different
approach, that is applicable for a general potential $V(x)$. Rather
than the constraint utilized in Pelinovsky {\it et al.} (2004),
Hwang {\it et al.} (2011) focused on the behaviour of the tails of
Bloch-wave packets. These tails are controlled by the coupling of
the wave envelope to the periodic Bloch mode at the band edge, an
effect that lies beyond all orders of the usual multiple-scale
expansion in powers of the wave amplitude. Hwang {\it et al.} (2011)
carried this expansion beyond all orders using techniques of
exponential asymptotics  (Yang \&\ Akylas 1997), for the case of
Bloch-wave packets whose envelope features a single hump. It turns
out that the tails of such wave packets decay as $x\to \pm \infty$,
and hence gap solitons arise, only for two specific locations of the
wave envelope relative to the underlying periodic potential. Both
these soliton families bifurcate from the linear (infinitesimal)
periodic Bloch mode at a band edge, and they coincide with the
on-site and off-site gap solitons found by Pelinovsky {\it et al.}
(2004) in the case of a symmetric periodic potential.

In the present work, making use of the asymptotic expressions
derived in Hwang {\it et al.} (2011) for the tails of a single-hump
Block-wave packet, we shall construct small-amplitude gap-soliton
families, in the form of bound states, that comprise two or more
such wave packets. In contrast to the so-called fundamental gap
solitons found in Pelinovsky {\it et al.} (2004) and Hwang {\it et
al.} (2011), these higher-order soliton families bifurcate at small
but finite amplitude close to a band edge, and they feature multiple
branches that do not connect to fundamental soliton families or band
edges.

In preparation for the ensuing analysis, we now summarize the main
results of the multiple-scale perturbation procedure for a
single-hump Bloch-wave packet (Pelinovsky {\it et al.} 2004; Hwang
{\it et al.} 2011). Close to a band edge $\mu=\mu_0$, say, gap
solitons are expected to be weakly nonlinear Bloch-wave packets. The
solution to equation \eqref{e:NLSamp} is expanded in powers of an
amplitude parameter, $0<\epsilon \ll 1$,
\[
\psi=\epsilon \psi_0+\epsilon^2 \psi_1+\epsilon^3 \psi_2+\cdots,
\label{e:psiexpansion}
\]
along with
\[
\mu=\mu_0+\eta \epsilon^2, \label{e:muexpansion}
\]
where $\eta=\pm 1$,
\[
\psi_0=A(X)p(x),
\]
$p(x)\equiv p(x;\mu_0)$, and $X=\epsilon x$ is the `slow' variable
of the envelope function $A(X)$. By imposing the appropriate solvability condition at $O(\epsilon^3)$, it turns out that $A(X)$ satisfies
the stationary NLS equation
\[
DA_{XX}+\eta A+\sigma\alpha A^3=0\,,\label{e:NLSenvelope}
\]
where
\[
D=\textstyle\frac12 \hspace{0.06cm} \d^2\hspace{-0.04cm}\mu/\d
\hspace{-0.01cm} k^2\,|_{\mu=\mu_0}\,, \qquad
\alpha={\int_0^{2\pi}p^4(x)\,\d x \bigg/ \int_0^{2\pi}p^2(x)\,\d
x}\,.  \label{e:diffraction}
\]
The well-known soliton solution of (\ref{e:NLSenvelope}) is
\[
A(X)=a \sech{X-X_0\over \beta}\,, \label{e:envelopeA}
\]
where
\[
a=\sqrt{2/\alpha}\,,\qquad \beta=\sqrt{|D|}\, ,  \label{e:defabeta}
\]
and $X_0=\epsilon x_0$ denotes the position of the peak of the
envelope. This solution exists provided $D\eta <0$ and $D\sigma >0$;
the first of these conditions requires $\mu$ to lie in the interior
of the band gap, while the second condition can be met in the
presence of self-focusing ($\sigma=1$) nonlinearity if $D>0$, or
self-defocusing ($\sigma=-1$) nonlinearity if $D<0$.

It is important to note that the envelope equation
\eqref{e:NLSenvelope} is translation-invariant, and $X_0$ is a free
parameter in the solution \eqref{e:envelopeA}. As a result, gap solitons of equation \eqref{e:NLSamp} could be obtained
regardless of the position of the envelope \eqref{e:envelopeA}
relative to the underlying periodic potential. This conclusion would
seem rather suspicious, though, given that the original amplitude
equation \eqref{e:NLSamp} is \emph{not} translation-invariant.

This issue was recognized by Pelinovsky {\it et al.} (2004), who
pointed out that locally confined solutions of \eqref{e:NLSamp} must
obey the constraint
\[
M(x_0)=\int_{-\infty}^\infty V'(x)\psi^2(x;x_0) \d x=0\,,
\label{e:Melnikov}
\]
which can be readily obtained by multiplying \eqref{e:NLSamp} with
$\psi_x$ and integrating with respect to $x$. Upon inserting the
perturbation solution \eqref{e:psiexpansion} and the potential
\eqref{e:potential} into \eqref{e:Melnikov}, this condition then
restricts the peak of the envelope to be at either a minimum
($x_0=0$) or a maximum ($x_0=\pi/2$) of the potential
\eqref{e:potential}, corresponding to on-site or off-site gap
solitons, respectively. It is also worth noting that the so-called
Melnikov function $M(x_0)$ in \eqref{e:Melnikov}, which depends on
the shift $x_0$ of the envelope relative to the potential $V(x)$, is
exponentially small with respect to $\epsilon$; hence, the
constraint (\ref{e:Melnikov}) for possible locations of gap solitons
hinges upon information beyond all orders of the two-scale expansion
\eqref{e:psiexpansion}.

\section{Nonlocal solitary waves}
\label{s:nonlocalwaves}

Another way of reconciling the two-scale expansion
\eqref{e:psiexpansion} with the fact that gap solitons can be placed
only at specific locations relative to the potential $V(x)$, is by
examining the behaviour of the tails of Bloch-wave packets near a
band edge. Assuming that they have infinitesimal amplitude, these
tails are governed by the linear version of equation
\eqref{e:NLSamp}. Hence, for $\mu$ inside a band gap and close to
the edge $\mu_0$, using the same notation as in
\eqref{e:muexpansion} and \eqref{e:diffraction}, the asymptotic representation of $\psi
(x)$ away from the solitary-wave core is, generically, a linear combination of two evanescent Bloch
modes
\[
\psi_{\pm}(x)=p(x)\exp\bigg\{\pm\bigg({\mu_0-\mu\over
D}\bigg)^{\frac12} x\bigg\}\,, \label{e:combination}
\]
to leading order in $|\mu-\mu_0|=\epsilon^2 \ll 1$. As expected, the slow exponential
growth/decay of these modes is consistent with the behaviour of the tails of the envelope
function $A(X)$ according to the steady NLS equation \eqref{e:NLSenvelope}.

Now, for the purpose of constructing a solitary-wave solution of
\eqref{e:NLSamp}, the left-hand tail must involve only $\psi_+(x)$,
$\psi(x)\sim \epsilon \hspace{0.04cm} C_+\psi_+ (x)$, so that
$\psi\to 0$ as $x\to -\infty$. For generic values of $C_+$, when one
integrates equation (\ref{e:NLSamp}) from this left-hand tail to the
right-hand side, the right-hand tail would involve both $\psi_-(x)$
and $\psi_+(x)$,
\[
\psi(x)\sim \epsilon\, C_-\psi_-(x) + E_+\psi_+(x) \qquad (x\gg
1)\,, \label{e:righthand}
\]
so $\psi(x)$ is not locally confined. However, the amplitude $E_+$
of the growing tail in (\ref{e:righthand}) is exponentially small
with respect to $\epsilon$ (see below), and hence this tail cannot
be captured by an expansion in powers of $\epsilon$, such as
\eqref{e:psiexpansion}. Only when $C_+$ takes certain special values
can the growing-tail amplitude $E_+$ vanish, resulting in a true
solitary wave solution. This restriction on $C_+$ is consistent with
the integral constraint \eqref{e:Melnikov} utilized by Pelinovsky
{\it et al.} (2004).

Hwang {\it et al.} (2011) computed the amplitude $E_+$ by carrying expansion
\eqref{e:psiexpansion} beyond all orders of $\epsilon$ for a general
periodic potential $V(x)$, utilizing an
exponential-asymptotics procedure in the wavenumber domain (Akylas
\&\ Yang 1995; Yang \&\ Akylas 1997). According to this revised
perturbation analysis, $E_+$, which indeed is exponentially small
with respect to $\epsilon$, does vanish for two specific positions
of the envelope \eqref{e:envelopeA} relative to the underlying
potential, thus furnishing two families of gap solitons that
bifurcate at the band edge. For the symmetric periodic potential
\eqref{e:potential}, in particular, $E_+$ vanishes when the peak of
the envelope is placed at $x_0=0$ or $x_0=\pi/2$, corresponding to
the on-site and off-site gap solitons, as was obtained earlier by
Pelinovsky {\it et al.} (2004).

As the details of the exponential-asymptotics procedure are rather
involved, we shall only quote the main results. When the peak of the
envelope \eqref{e:envelopeA} is not at a minimum or maximum of the
potential ($x_0\ne 0,\pi/2$), the resulting `solitary' waves are
nonlocal owing to the presence of both growing and decaying
solutions \eqref{e:combination} in at least one of the tails.
Specifically, assuming that the left-hand tail is locally confined
in accordance with \eqref{e:psiexpansion} and \eqref{e:envelopeA},
\begin{subequations}
\label{e:updownstream}
\begin{gather}
\psi\sim 2\epsilon \hspace{0.02cm} a p(x)\e^{\epsilon
(x-x_0)/\beta}\qquad (x \to -\infty)\,, \label{e:upstream}
\end{gather}
the right-hand tail features a growing component of exponentially-small amplitude
with respect to $\epsilon$, in addition to a decaying component analogous to \eqref{e:upstream}:
\begin{align}
\psi\sim  & \ 2\epsilon \hspace{0.02cm}  a p(x)  \e^{-\epsilon (x-x_0)/\beta} \nonumber\\
&+4\pi C{\beta^3\over \epsilon^3}
\e^{-\pi\beta/\epsilon}\sin2x_0\,p(x)\e^{\epsilon (x-x_0)/\beta}
\qquad (x-x_0 \gg 1/\epsilon)\,.\label{e:downstream}
\end{align}
\end{subequations}
Note that the amplitude of the growing tail in \eqref{e:downstream}
is proportional to a constant $C$. As explained in Hwang {\it et
al.} (2011), this constant is determined from solving a certain
recurrence relation that contains information beyond all orders of
expansion \eqref{e:psiexpansion}. In particular, $C>0$ for the
sinusoidal potential (\ref{e:potential}). As expected, the growing
component of the right-hand tail \eqref{e:downstream} is absent when
$x_0=0,\pi/2$.

By utilizing the symmetries of equation \eqref{e:NLSamp} as noted
below, it is straightforward to deduce from \eqref{e:updownstream}
the asymptotic behaviour of nonlocal solitary waves whose right-hand
tail is locally confined,
\begin{subequations}
\label{e:downupstream}
\begin{gather}
\psi\sim 2\epsilon a p(x)\e^{-\epsilon (x-x_0)/\beta}\qquad (x \to
+\infty)\,, \label{e:downstreamb}
\end{gather}
but the left-hand tail comprises a growing and a decaying component:
\begin{align}
\psi\sim  & \ 2\epsilon  a p(x)  \e^{\epsilon (x-x_0)/\beta} \nonumber\\
& -4\pi C{\beta^3\over \epsilon^3}
\e^{-\pi\beta/\epsilon}\sin2x_0\,p(x)\e^{-\epsilon (x-x_0)/\beta}
\qquad (x-x_0 \ll -1/\epsilon)\,.\label{e:upstreamb}
\end{align}
\end{subequations}
For the symmetric potential (\ref{e:potential}), equation
(\ref{e:NLSamp}) is invariant with respect to reflection in $x$
$(x\rightarrow -x)$, and the Bloch wave $p(x)$ at a band edge is
either symmetric or antisymmetric in $x$, $p(-x)=\pm p(x)$. Since
(\ref{e:NLSamp}) is also invariant with respect to $\psi\to -\psi$,
the reflected solution $\psi$ may thus be written as
(\ref{e:downupstream}) for both symmetric and antisymmetric $p(x)$.
The asymptotic expressions \eqref{e:updownstream} and
\eqref{e:downupstream} for the tails of nonlocal solitary waves are
key to constructing new families of locally confined solutions, in
the form of bound states, as discussed below.

\section{Bound states}
\label{s:boundstates}

As indicated above, gap solitons in the form of Bloch-wave packets
with the `sech' envelope \eqref{e:envelopeA} arise only when the
peak of the envelope is at a minimum ($x_0=0$) or a maximum
($x_0=\pi/2$) of the potential \eqref{e:potential}. Apart from these
fundamental soliton states, it is possible, however, to construct other gap-soliton families by piecing together two or
more of the nonlocal solitary waves found in
\S\ref{s:nonlocalwaves}. The overall strategy for determining these
so-called bound states is similar to that followed in Yang \&\
Akylas (1997) for constructing multi-packet solitary-wave solutions
of the fifth-order KdV equation. Here, we shall work out the details
of finding bound states involving two Bloch-wave packets.

\subsection{Matching of tails}
\label{s:matching}

Consider two nonlocal solitary-wave solutions, $\psi^+(x)$ and
$\psi^-(x)$, whose `sech'-profile envelope functions are centred at
$x_0^+$ and $-x_0^-$, respectively, with $x_0^\pm>0$. In addition,
let $\psi^\pm(x)\to 0$ as $x\to\pm \infty$, so the right-hand
tail of $\psi^-(x)$ and the left-hand tail of $\psi^+(x)$ are
nonlocal. Let us define the separation between the solitary-wave
cores of the two constituent nonlocal waves as
\[  \label{Sdef}
S \equiv x_0^+ + x_0^-.
\]
Then, assuming $S\gg 1/\epsilon$, \eqref{e:downstream} and
\eqref{e:upstreamb} are legitimate asymptotic expressions for these
tails in the `overlap' region $-x_0^-\ll x \ll x_0^+$. Specifically,
according to \eqref{e:downstream}, the right-hand tail of
$\psi^-(x)$ is given by
\[
\psi^-\sim  \pm 2\epsilon  a \hspace{0.05cm}  \e^{-\epsilon
(x+x_0^-)/\beta} p(x) \mp 4\pi C{\beta^3\over \epsilon^3}
\e^{-\pi\beta/\epsilon}\sin2x_0^-\,\e^{\epsilon (x+x_0^-)/\beta}p(x)
\,,\label{e:psiminus}
\]
and, according to \eqref{e:upstreamb}, the left-hand tail of
$\psi^+(x)$ reads
\[
\psi^+\sim  2\epsilon  a \hspace{0.05cm}  \e^{\epsilon
(x-x_0^+)/\beta} p(x) -  4\pi C{\beta^3\over \epsilon^3}
\e^{-\pi\beta/\epsilon}\sin2x_0^+\,\e^{-\epsilon
(x-x_0^+)/\beta}p(x) \,.\label{e:psiplus}
\]
Note that the upper sign in \eqref{e:psiminus} corresponds to the
case where the envelopes of the two solitary waves have the same
polarity (sign), while the lower sign pertains to the case of
opposite polarity.

Now, in order for these two nonlocal solitary waves to form a bound state, their growing
and decaying tail components must match smoothly in the overlap region between the
two main cores. Based on \eqref{e:psiminus} and \eqref{e:psiplus}, this requires
\[
\sin 2x_0^-=\sin 2x_0^+=
\mp {a\over 2\pi C}{\epsilon^4\over \beta^3}
\e^{\pi\beta/\epsilon}\,\e^{-\epsilon S/\beta}\,,
\label{e:matchingcond}
\]
and, for this matching condition to be met, it is
necessary that
\[
{a\over 2\pi C}{\epsilon^4\over \beta^3}
\e^{\pi\beta/\epsilon}\,\e^{-\epsilon S/\beta} \le 1\,.
\label{e:constraint}
\]
Clearly, owing to the growing exponential $\e^{\pi\beta/\epsilon}$ on the
left-hand side of \eqref{e:constraint}, this constraint cannot be
satisfied for $\epsilon\to 0$ when $S$ is finite. As a result,
solution families of bound states are expected to bifurcate at
a finite amplitude $\epsilon=\epsilon_c$, say, depending on
the separation distance $S$. Below, we shall verify this claim and
compute $\epsilon_c$ for the various solution
branches.

\subsection{Solution branches}
\label{s:branches}

Attention is now focused on the matching condition \eqref{e:matchingcond}, in order to
delineate the solution branches of bound states that bifurcate close to a band edge.

Consider first the case where the two nonlocal solitary waves have
envelopes with the same polarity. Then, the upper sign applies in
\eqref{e:matchingcond}, and $\sin2x_0^-=\sin2x_0^+ <0$. This
condition can be satisfied in two ways:
\begin{subequations}
\label{e:cond1}
\begin{align}
{\rm (i)} \quad  & x_0^-=(m+{\textstyle\frac12})\pi +d\,, \qquad
x_0^+=(n-{\textstyle\frac12)}\pi +d\,;
\label{e:cond1a} \\
{\rm (ii)} \quad & x_0^-=(m+{\textstyle\frac12})\pi +d\,, \qquad  x_0^+=n\pi -d\,,
\label{e:cond1b}
\end{align}
\end{subequations}
where $m$, $n$ are integers and $0<d<\pi/2$. Accordingly, the
solitary-wave separation \eqref{Sdef} corresponding to these two possibilities
is
\[
{\rm (i)} \quad S=N\pi+2d\,; \qquad {\rm (ii)} \quad
S=(N+{\textstyle\frac12})\pi\,, \label{e:conditionS}
\]
where $N=m+n$. Moreover, the matching condition \eqref{e:matchingcond} reads
\[
\sin 2d={a\over 2\pi C}{\epsilon^4\over \beta^3}
\e^{\pi\beta/\epsilon}\,\e^{-\epsilon S/\beta}\,,
\label{e:conditionsine}
\]
with $S$ given by (i) and (ii) in \eqref{e:conditionS}.

It is convenient to label families of bound states by the integer
$N$ which, in view of \eqref{e:conditionS}, controls the separation
between the two nonlocal solitary waves that form the bound states.
As noted earlier, the present theory assumes that these solitary
waves are well-separated, namely, $S\gg 1/\epsilon$, which in turn
requires $N\gg 1/\epsilon$; the validity of this condition will be
verified later in this section.

In preparation for tracing the solution branches of the family of
bound states corresponding to a given $N$,
we write
\[
d=\pi/4 +\delta,
\]
with $-\pi/4 <\delta < \pi/4$, so that from \eqref{e:conditionS}, we
have
\[
{\rm (i)} \quad S=S_0+2\delta\,; \qquad {\rm (ii)} \quad S=S_0\,,
\label{e:conditionS0}
\]
where $S_0=(N+\frac12)\pi$. The matching condition
\eqref{e:conditionsine}, with $S$ given by (i) and (ii) above, then leads
to the following two conditions, respectively,
\begin{subequations}
\label{e:conditioncosine}
\begin{align}
& \cos 2\delta=W(\epsilon, N) \,\e^{-2\epsilon \delta/\beta}\,,
\label{e:conditioncosinea}\\
& \cos 2\delta=W(\epsilon, N)\,,\label{e:conditioncosineb}
\end{align}
\end{subequations}
where
\[
W(\epsilon, N)={a\over 2\pi C}{\epsilon^4\over \beta^3}
\e^{\pi\beta/\epsilon}\e^{-\epsilon S_0/\beta}\,. \label{e:defW}
\]
Based on equations \eqref{e:conditioncosine}, given $\epsilon$, one
may find the values of $\delta$, and from \eqref{e:conditionS0} the
corresponding separations $S$, for which bound states are possible.

\begin{figure}[tb!]
\centerline{
\includegraphics[width=0.555\textwidth]{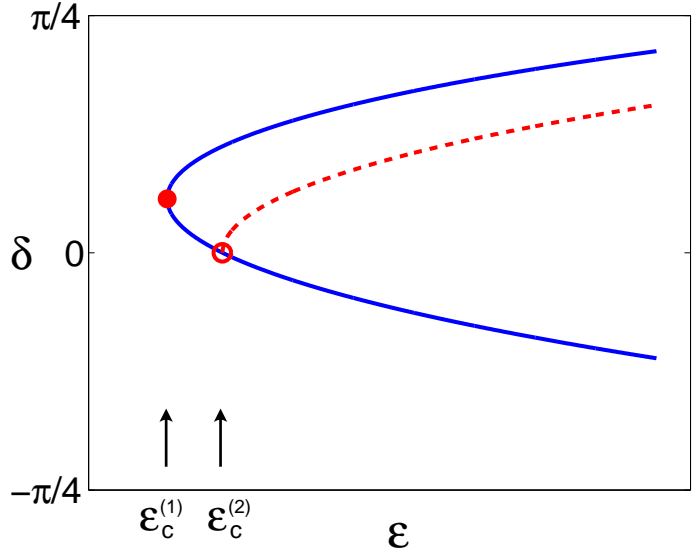}
}
\smallskip
\caption{Schematic diagram for the three branches of the solution
curve $\delta(\epsilon)$. Two branches (solid blue) are obtained
from solving \eqref{e:conditioncosinea} and one branch (dashed red)
from solving \eqref{e:conditioncosineb}. The bifurcation point
$\epsilon_c^{(1)}$ with $\delta_c^{(1)}>0$ (solid dot) is found by
solving \eqref{e:eqepsilonc}, and the bifurcation point
$\epsilon_c^{(2)}$ with $\delta_c^{(2)}=0$ (open dot) is found by
solving \eqref{e:eqepsiloncb}. (Online version in colour.)}
\label{f:delta_curve}
\end{figure}

Specifically, from equation \eqref{e:conditioncosinea}, it is easy
to see that two values of $\delta$ arise for each $\epsilon$ above a
certain threshold, $\epsilon_c^{(1)}$, which is the
bifurcation point of the bound states obeying
\eqref{e:conditioncosinea}. This threshold is associated with a double root,
$\delta=\delta_c^{(1)}$, of equation \eqref{e:conditioncosinea},
where
\[
\left. {\partial \over \partial\delta} \left( \cos
2\delta-W(\epsilon_c^{(1)}, N)
\,\e^{-2\epsilon_c^{(1)}\delta/\beta}\right)\right|_{\delta=\delta_c^{(1)}}=0\,.
\label{e:deltac}
\]
From this equation and (\ref{e:conditioncosinea}), it follows that
\begin{equation}
\delta_c^{(1)}=\frac12 \tan^{-1}{\epsilon_c^{(1)}/\beta}.
\end{equation}
Inserting this result into \eqref{e:conditioncosinea},
$\epsilon_c^{(1)}$ then is found by solving
\[
\cos\big(\tan^{-1}{\epsilon_c^{(1)}\over
\beta}\big)=W(\epsilon_c^{(1)}, N)
\exp\bigg\{-{\epsilon_c^{(1)}\over
\beta}\tan^{-1}{\epsilon_c^{(1)}\over \beta}\bigg\}\,.
\label{e:eqepsilonc}
\]
\indent{For $N\gg 1$, the two solution branches of
$\delta(\epsilon)$, obtained from \eqref{e:conditioncosinea}
for $\epsilon > \epsilon_c^{(1)}$, are plotted schematically in
figure \ref{f:delta_curve}. The bifurcation point
$\epsilon_c^{(1)}$ is the turning point of this double-branch curve,
and $\delta_c^{(1)}>0$. For $\epsilon \gg
\epsilon_c^{(1)}$, it is clear from \eqref{e:conditioncosinea} that
$\cos 2\delta \to 0$, and hence the two branches approach
$\delta=\pm \pi/4$, so $d=\pi/4 +\delta\to \pi/2, 0$. Therefore, in
this limit, in view of \eqref{e:cond1a}, the two solution branches
that bifurcate at $\epsilon_c^{(1)}$ approach limiting
configurations of bound states, with both solitary waves being
on-site on one branch (as $\delta\to \pi/4$) and off-site on the
other branch (as $\delta\to -\pi/4$). In other words, as the
solution curve in figure \ref{f:delta_curve} is traced from one
branch to the other, the two solitary waves making up the bound
state transform simultaneously from on-site to off-site. In this
transition, $\delta$ decreases from $\pi/4$ to $-\pi/4$, and
according to (i) in (\ref{e:conditionS0}), the separation distance
$S$ of the two solitary waves decreases from $(N+1)\pi$ to $N\pi$.
This transition behaviour will be verified numerically in \S6; see
figure \ref{f:same_polarity}.}

Turning next to the second possibility of bound states in \eqref{e:conditionS0},
equation \eqref{e:conditioncosineb} furnishes two values of $\delta$,
\[
\delta=\pm\frac12 \cos^{-1} W(\epsilon, N),    \label{e:eqdelta}
\]
for each $\epsilon$, as long as $W(\epsilon, N)\le 1$. Hence,
$\epsilon$ must exceed a certain threshold, $\epsilon \ge
\epsilon_c^{(2)}$, where $W(\epsilon_c^{(2)}, N)=1$. In view of
\eqref{e:defW}, this bifurcation point is determined from the equation
\[
{\pi\beta \over \epsilon_c^{(2)}}-{\epsilon_c^{(2)} S_0\over \beta}
=\ln\bigg[ {2\pi C\over a}\cdot {\beta^3 \over \epsilon_c^{(2)^4}} \bigg]\,.
\label{e:eqepsiloncb}
\]
Here, the two values of $\delta$ in \eqref{e:eqdelta}
actually correspond to equivalent bound-state configurations (under
mirror reflection of the $x$-axis), as can be easily verified from
\eqref{e:cond1b} with $d=\frac14 \pi +\delta$. Therefore,
\eqref{e:conditioncosineb} describes only one solution branch. For
$\epsilon \gg \epsilon_c^{(2)}$, in particular, $\delta\to \pm
\pi/4$, and this solution branch approaches a limiting bound-state
configuration that, according to \eqref{e:cond1b}, involves an
on-site and an off-site solitary wave.

We remark that the double root $\delta=\delta_c^{(2)}=0$ of
\eqref{e:conditioncosineb} for $\epsilon=\epsilon_c^{(2)}$ is also a
solution of \eqref{e:conditioncosinea} for the same value of
$\epsilon$. Hence, the solution branch \eqref{e:conditioncosineb}
bifurcates from the point $\delta=0$ on the lower branch of the
solutions \eqref{e:conditioncosinea}, which themselves bifurcate at
$\epsilon_c^{(1)}$, as illustrated schematically in figure
\ref{f:delta_curve}. As a consequence, the bound-state family
corresponding to a certain $N$ comprises three distinct solution
branches which are connected with each other. The bifurcation
points, $\epsilon_c^{(1)}$ and $\epsilon_c^{(2)}$, do not coincide;
but when $N$ is large, they are close to each other. Indeed, in the
limit $N\gg 1$, it can be readily shown, using
\eqref{e:conditioncosine}, \eqref{e:defW} and \eqref{e:eqepsiloncb},
that $\epsilon_c^{(1)}\sim \epsilon_c^{(2)}\sim \beta N^{-1/2}$.
Thus, for large solitary-wave separation ($N\gg 1$), bound-state
families bifurcate near a band edge, but never exactly at the band
edge itself. Moreover, this scaling of $\epsilon_c^{(1)}$ and
$\epsilon_c^{(2)}$ in terms of $N$ is consistent with the condition
$N\gg 1/\epsilon$ imposed earlier for the purpose of matching the
tails (\ref{e:psiminus}) and (\ref{e:psiplus}) of nonlocal solitary
waves forming a bound state.

We now consider the case when the two nonlocal solitary waves
participating in a bound state have envelopes with opposite
polarity, and the lower sign in \eqref{e:matchingcond} applies.
Hence, $\sin 2x_0^+=\sin 2x_0^- >0$. This condition can be satisfied
in two ways:
\begin{subequations}
\label{e:cond2}
\begin{align}
{\rm (i)} \quad  & x_0^-=m\pi +d\,, \qquad    x_0^+=n\pi +d\,;
\label{e:cond2a} \\
{\rm (ii)} \quad & x_0^-=m\pi +d\,, \qquad  x_0^+=(n
+{\textstyle\frac12})\pi -d\,, \label{e:cond2b}
\end{align}
\end{subequations}
for some integers $m,n$ and $0<d<\pi/2$. As a result, the
expressions \eqref{e:conditionS} for the solitary-wave separation
$S=x_0^+ + x_0^-$ are also valid here, so the matching condition
\eqref{e:matchingcond} gives rise to the same equations
\eqref{e:conditioncosine} that describe the solution branches
$\delta=\delta(\epsilon)$ of bound states belonging to the family
labelled by $N=m+n$, where $d=\pi/4 +\delta$ as before.
Specifically, for a given $N$, the three solution branches found
earlier arise in this instance as well: two of these branches
bifurcate at $\epsilon=\epsilon_c^{(1)}$ and, in the limit $\epsilon
\gg \epsilon_c^{(1)}$, represent bound states with both solitary
waves being on-site ($\delta\to -\pi/4$) or off-site ($\delta\to
\pi/4$). Marching along this solution curve from the on-site branch
to the off-site branch, $\delta$ increases from $-\pi/4$ to $\pi/4$,
and, in view of (\ref{e:cond2a}), the separation distance $S$ of the
two nonlocal solitary waves increases from $N\pi$ to $(N+1)\pi$.
(This transition behaviour is also brought out by the numerical
results presented in \S6; see figure \ref{f:opposite_polarity}.)
This transition contrasts with the case of same envelope polarity
examined earlier, where the separation distance $S$ of the two
nonlocal solitary waves decreases from $(N+1)\pi$ to $N\pi$ when
marching along the solution curve from the on-site branch to the
off-site branch.

Finally, according to \eqref{e:conditioncosineb},
\eqref{e:eqdelta} and \eqref{e:eqepsiloncb},
the third solution branch bifurcates at
$\epsilon=\epsilon_c^{(2)}$ and, for $\epsilon \gg
\epsilon_c^{(2)}$, represents bound states with one solitary wave
on-site and the other off-site. Moreover, it follows from \eqref{e:cond2}
that, in the course of continuation of the bound-state solution from the
on-site branch to this mixed-site branch, one of the two on-site
solitary waves remains on-site and does not move, while the other
on-site solitary wave moves away and becomes off-site.

\section{Power curves}
\label{s:powercurves}

In applications, gap-soliton solution branches are often described through
\[
P=\int_{-\infty}^{\infty} \psi^2 \d x\,,
\label{e:power}
\]
which is commonly referred to as the soliton power. This quantity is
also key to understanding the stability properties of gap solitons
(Vakhitov \& Kolokolov 1973; Shi {\it et al.} 2008; Yang 2010).
Here, we shall derive an analytical expression for the power of
small-amplitude bound states near band edges, and trace the solution
branches found earlier in terms of $P$. The theoretical predictions
will be compared against numerical results in \S\ref{s:numerical}.

By virtue of perfect matching of the tails \eqref{e:psiminus} and
\eqref{e:psiplus} in the overlap region $-x_0^-\ll x \ll x_0^+$, one
may write the following uniformly valid approximation to bound
states involving two nonlocal solitary waves (to the leading order):
\[
\psi\sim \epsilon a \sech\{\epsilon(x+x_0^-)/\beta\}p(x)\pm
\epsilon a \sech\{\epsilon(x-x_0^+)/\beta\}p(x)\,;
\label{e:psiapproxi}
\]
here again the upper (lower) sign corresponds to the case where the
envelopes of the solitary waves have the same (opposite) polarity.
Inserting \eqref{e:psiapproxi} into \eqref{e:power}, the power
associated with a bound state may then be approximated as
\[
P\sim I^++I^- \pm I\,,
\label{e:powerapproxi0}
\]
where
\begin{gather}
I^{\pm} =\epsilon^2 a^2 \int_{-\infty}^{\infty}
p^2(x)\sech^2\{\epsilon(x\mp x_0^{\pm})/\beta\}\d x,
\label{e:defIpm}
\\
I =2\epsilon^2 a^2 \int_{-\infty}^{\infty} p^2(x)\sech\{\epsilon(x+x_0^{-})/\beta\}
\sech\{\epsilon(x-x_0^{+})/\beta\} \d x\,.
\label{e:defI}
\end{gather}

The integrals above can be evaluated by substituting the Fourier series
\[  \label{e:p2fourier}
p^2(x)=q_0+\sum_{m=1}^{\infty} q_m \cos 2mx,
\]
for the even $\pi$-periodic function $p^2(x)$, into \eqref{e:defIpm}
and \eqref{e:defI} and then integrating term-by-term. Specifically,
in the limit $\epsilon \ll 1$, $\epsilon S \gg 1$, we find
\begin{gather}
I^{\pm} \sim 2a^2 \epsilon\beta  q_0 + 4\pi a^2\beta^2
q_1\cos2x_0^{\pm} \hspace{0.04cm}  \e^{-\pi\beta/\epsilon}
+\cdots\,, \label{e:defIpmapproxi}\\   \hspace{-2.65cm} I \sim 8a^2
\epsilon^2 q_0 S\e^{-\epsilon S/\beta}+\cdots\,.
\label{e:defIapproxi}
\end{gather}
From \eqref{e:conditionsine}, however, it is clear that $\epsilon^2
S\e^{-\epsilon S/\beta}\gg \e^{-\pi \beta/\epsilon}$; hence, the
second term in \eqref{e:defIpmapproxi} is sub-dominant in comparison
to \eqref{e:defIapproxi}, and \eqref{e:powerapproxi0} yields the
following asymptotic expression for $P$,
\[
P\sim 4a^2\epsilon q_0 \big( \beta \pm 2 \epsilon S \e^{-\epsilon
S/\beta}\big) \qquad (\epsilon S\gg 1\,,  \  \epsilon\ll 1)\,,
\label{e:powerapproxi}
\]
with
\[
q_0=\frac1\pi \int_0^\pi p^2(x) \d x\,.
\]

Consider first the case of bound states comprising two solitary
waves with envelopes of the same polarity, where the upper sign in
\eqref{e:powerapproxi} applies. Corresponding to the first
possibility for $S$ in \eqref{e:conditionS0}, there are two
bound-state solution branches, $\delta=\delta(\epsilon)$, governed
by \eqref{e:conditioncosinea}, and the associated power according to
\eqref{e:powerapproxi} is
\[
P\sim 4a^2 \epsilon q_0 \bigg\{ \beta+ 2\epsilon (S_0+2\delta)
\e^{-\epsilon (S_0+2\delta)/\beta}\bigg\}\,,\label{e:powerapproxi1a}
\]
where $S_0=(N+{\textstyle\frac12})\pi$ as before. On the other hand,
corresponding to the second possibility for $S$ in
\eqref{e:conditionS0}, there is only one solution branch, given by
\eqref{e:eqdelta}, and its power is
\[
P\sim 4a^2 \epsilon q_0 \bigg\{ \beta+ 2\epsilon S_0 \e^{-\epsilon
S_0/\beta}\bigg\} \,.  \label{e:powerapproxi1b}
\]
In both \eqref{e:powerapproxi1a} and \eqref{e:powerapproxi1b}, the
second term in the brackets derives from the interaction of the
tails of the nonlocal solitary waves. Moreover, it should be kept in mind that
these expressions are valid when $\epsilon\ll 1$ and $\epsilon
S\gg 1$; i.e., close to the band edge and for wide separation
of the two solitary waves forming a bound state.

Based on the asymptotic formulae \eqref{e:powerapproxi1a} and
\eqref{e:powerapproxi1b}, it is possible to deduce the qualitative
behaviour of the power curves.  We recall that, for bound states of
solitary waves with the same envelope polarity, $\delta$ decreases
from $\pi/4$  to $-\pi/4$ as the solution curve
\eqref{e:conditioncosinea} is traversed from the on-site to the
off-site branch.  Since, for given $\epsilon$, the power
\eqref{e:powerapproxi1a} is a decreasing function of $\delta$ (for
$\epsilon S>\beta$), the power on the off-site branch is then always
higher than the power on the on-site branch.  Similarly, one can see
that the power \eqref{e:powerapproxi1b} of the mixed-site solution
branch \eqref{e:conditioncosineb} always lies between the powers of
the on-site and off-site braches.  Hence, the power curve associated
with a bound-state solution family involving solitary waves of the
same polarity, is such that the on-site branch is always the
lower-power branch, the mixed-site branch lies in the middle and the
off-site branch is the higher-power branch.  Also, in the
$(\epsilon,\delta)$ plane, the middle solution branch bifurcates
from the lower (off-site) branch (see figure 1); thus, on the power
curve, the intermediate-power (mixed-site) branch bifurcates from
the higher-power (off-site) branch.  These general features of the
power curves will be verified numerically in $\mathsection$6 (see
figures 2 and 5).


Next, in the case of bound states with nonlocal solitary waves having envelopes of
opposite polarity, where the lower sign in \eqref{e:powerapproxi}
applies, a similar calculation based on \eqref{e:cond2}, along with
\eqref{e:conditionS0}, yields
\[
P\sim 4a^2 \epsilon q_0 \bigg\{ \beta- 2\epsilon (S_0+2\delta)
\e^{-\epsilon (S_0+2\delta)/\beta}\bigg\}\,,\label{e:powerapproxi2a}
\]
for the two solution branches bifurcating at $\epsilon_c^{(1)}$, and
\[
P\sim 4a^2 \epsilon q_0 \bigg\{ \beta-2\epsilon S_0 \e^{-\epsilon
S_0/\beta}\bigg\} \,,   \label{e:powerapproxi2b}
\]
for the single solution branch bifurcating at $\epsilon_c^{(2)}$.
Based on these power formulae, it is again possible to show that, on
the associated power curve, the on-site branch is the lower-power branch,
the mixed-site branch is the immediate-power branch, and the off-site branch
is the higher-power branch, just as in the case of same envelope polarity
above. However, the intermediate-power branch now bifurcates from the
lower-power branch, which is the opposite of the conclusion reached
earlier in the same-envelope-polarity case. These theoretical
predictions are confirmed by numerical results below (see figure
\ref{f:opposite_polarity}).

\section{Numerical results}
\label{s:numerical}

We now turn to a numerical investigation of bound states in order to
make a comparison of numerical results against the predictions of
the asymptotic theory discussed earlier. To this end,
equation~\eqref{e:NLSamp} is solved numerically by the
Newton-conjugate-gradient method (Yang 2010), using the sinusoidal
potential (\ref{e:potential}) with  $V_0=6$. For this value of the
potential depth, when $\sigma=1$ (self-focusing nonlinearity), gap
solitons bifurcate from the lower edge $\mu_0=2.061318$ of the
first Bloch band, where the diffraction coefficient $D$ is positive,
and the parameters \eqref{e:defabeta} of the envelope soliton
\eqref{e:envelopeA} take the values
\[
a=1.7146\,, \qquad \beta=\sqrt{D}=0.6594\,.
\]
Moreover, the constant $C$ in the tail expressions
\eqref{e:downstream} and \eqref{e:upstreamb} is found by solving a
certain recurrence relation as discussed in Hwang {\it et al.}
(2011): $C=1.307$. On the other hand, when $\sigma=-1$
(self-defocusing nonlinearity), gap solitons bifurcate from the
upper edge $\mu_0=2.266735$ of the first Bloch band, where the
diffraction coefficient is negative, and the soliton parameter values are
\[
a=1.681\,, \qquad \beta=\sqrt{|D|}=0.7669\,,
\]
with $C=3.0133$.

\begin{figure}[t!]
\centerline{\includegraphics[width=0.9\textwidth]{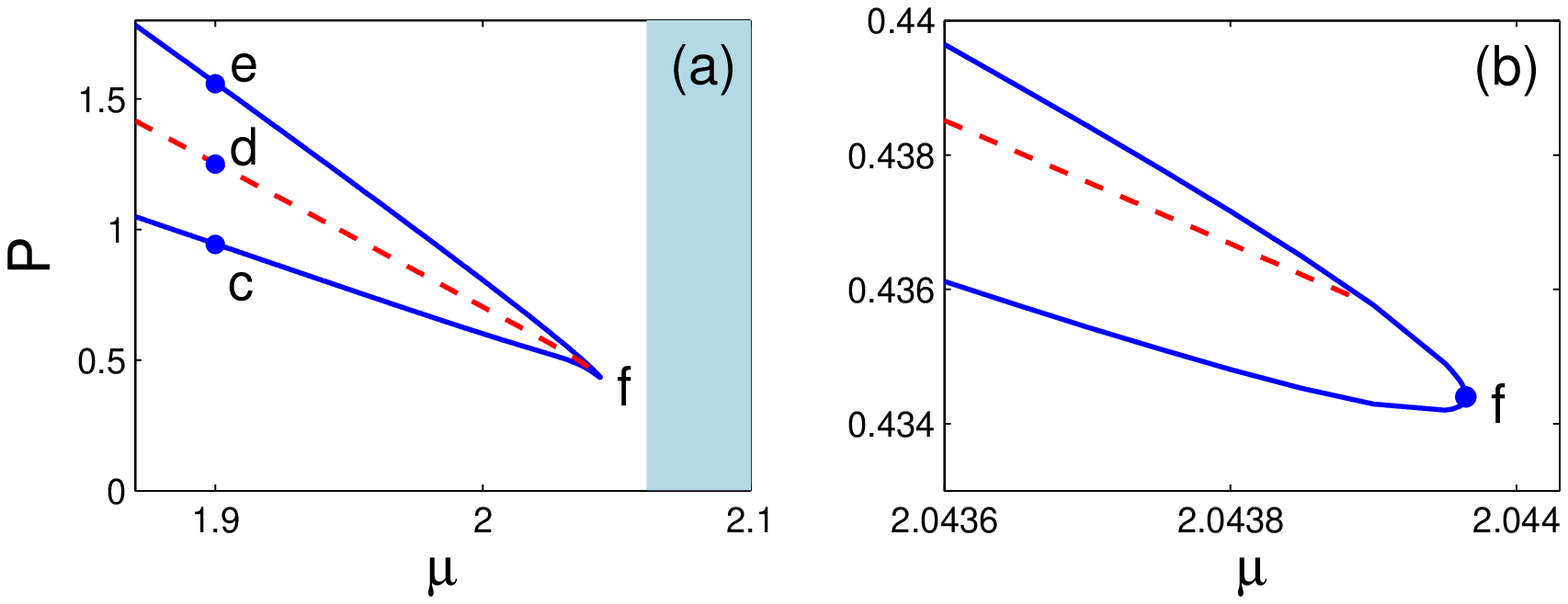}}

\vspace{0.8cm}
\centerline{\includegraphics[width=0.9\textwidth]{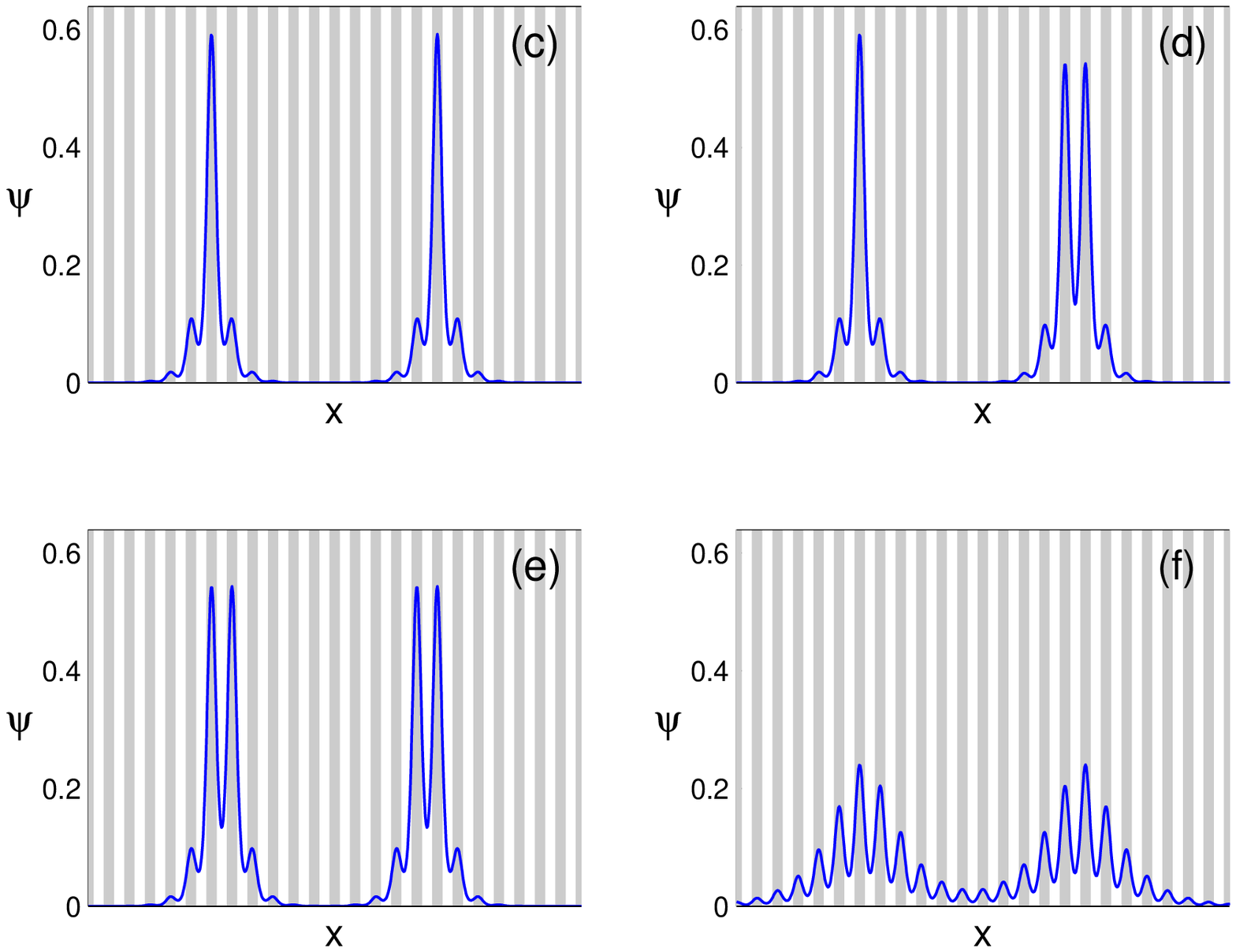}}

\caption{({\it a}) Numerical power curve for bound states involving
two solitary waves of the same envelope polarity with $N=10$ and
$\sigma=1$ (self-focusing nonlinearity); the shaded region is the
first Bloch band. ({\it b}) Amplification of ({\it a}) near the
bifurcation point. ({\it c--f}) Soliton profiles at points of the
power curve marked by the same letters in ({\it a}).  Shaded stripes
represent lattice sites (locations of low potentials); stripe
separation is equal to the potential period, $\pi$. }
\label{f:same_polarity}
\end{figure}

We begin with a few qualitative comparisons between the analysis and
numerics. For this purpose, we assume self-focusing nonlinearity
($\sigma=1$) and consider bound states involving solitary waves with
envelopes of the same polarity for $N=10$. The numerical results for
this family of bound states are displayed in figure
\ref{f:same_polarity}. From the power curve in figure
\ref{f:same_polarity}{\it a}, it is seen that this family indeed
exhibits three distinct solution branches that bifurcate near the
band edge, as predicted by the theory. On the lower-power branch,
the bound state at point c comprises two on-site gap solitons which
are separated by $(N+1)\pi=11\pi$ (see figure
\ref{f:same_polarity}{\it c}); on the higher-power branch, the bound
state at point e comprises two off-site gap solitons which are
separated by $N\pi=10\pi$ (see figure \ref{f:same_polarity}{\it e}).
Thus, as one moves from the lower branch to the upper branch along
the power curve, the two solitary waves forming a bound state
transform from on-site to off-site, and their separation decreases by
one lattice period ($\pi$). On the middle branch, however, the bound
state at point d comprises an on-site and an off-site gap soliton
which are separated by $(N+1/2)\pi=10.5\pi$ (see figure
\ref{f:same_polarity}{\it d}), confirming that this is the
mixed-site branch. At the bifurcation point near the band edge (tip
of the power curve), the bound-state profile is displayed in figure
\ref{f:same_polarity}{\it f}. This bound state comprises two
low-amplitude Bloch-wave packets which are well separated, as
assumed in our asymptotic analysis. These features of the power
curve and the corresponding bound-state profiles are in perfect
qualitative agreement with the asymptotic theory. Furthermore, from
the amplification of the power curve near the bifurcation point,
shown in figure \ref{f:same_polarity}{\it b}, it is clear that the
middle branch bifurcates from the higher-power branch, again in agreement
with the previous analysis in the case of bound states with solitary waves of the same
envelope polarity.

\begin{figure}[t!]
\centerline{
\includegraphics[width=0.42\textwidth]{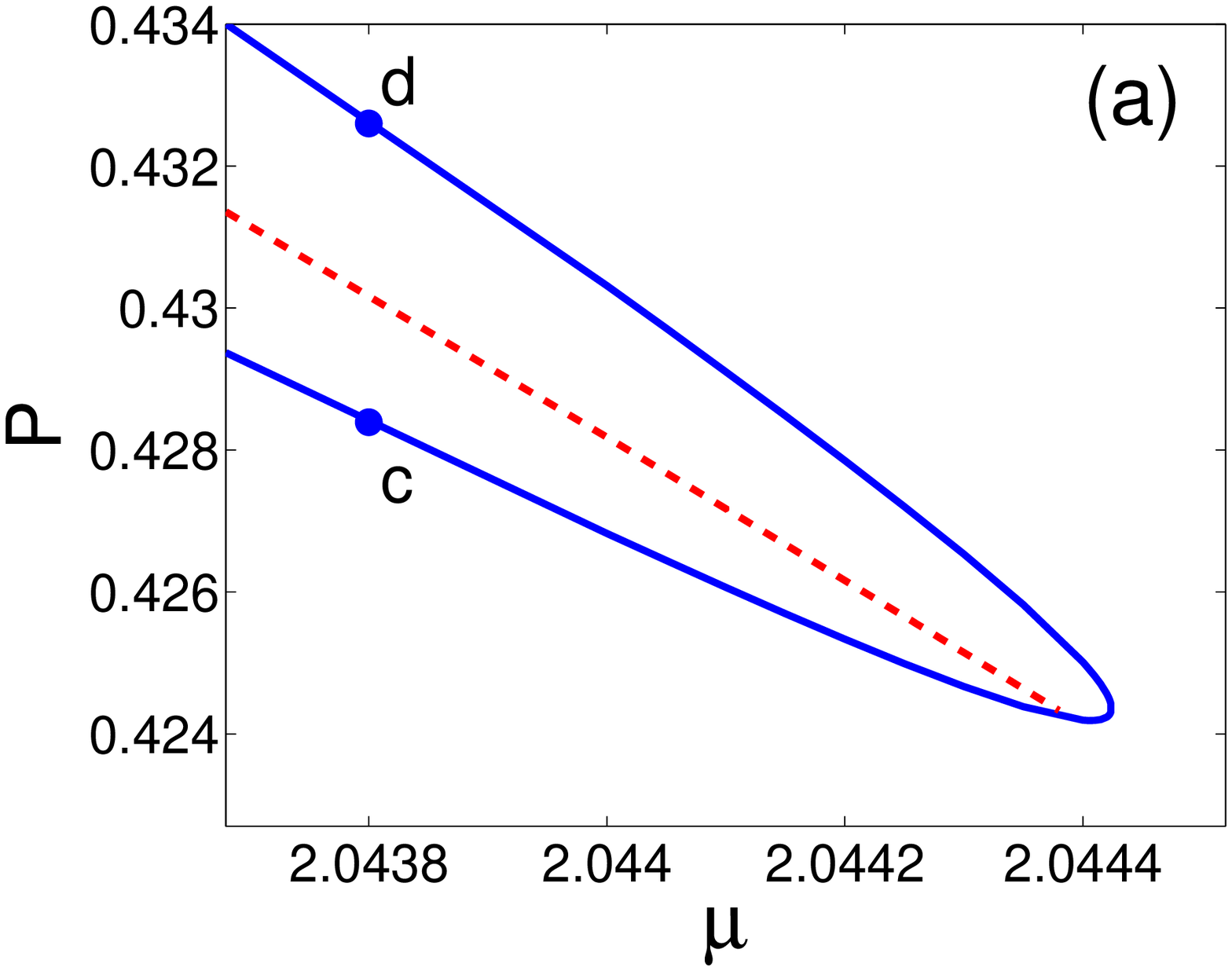} \hspace{0.3cm}
\includegraphics[width=0.445\textwidth]{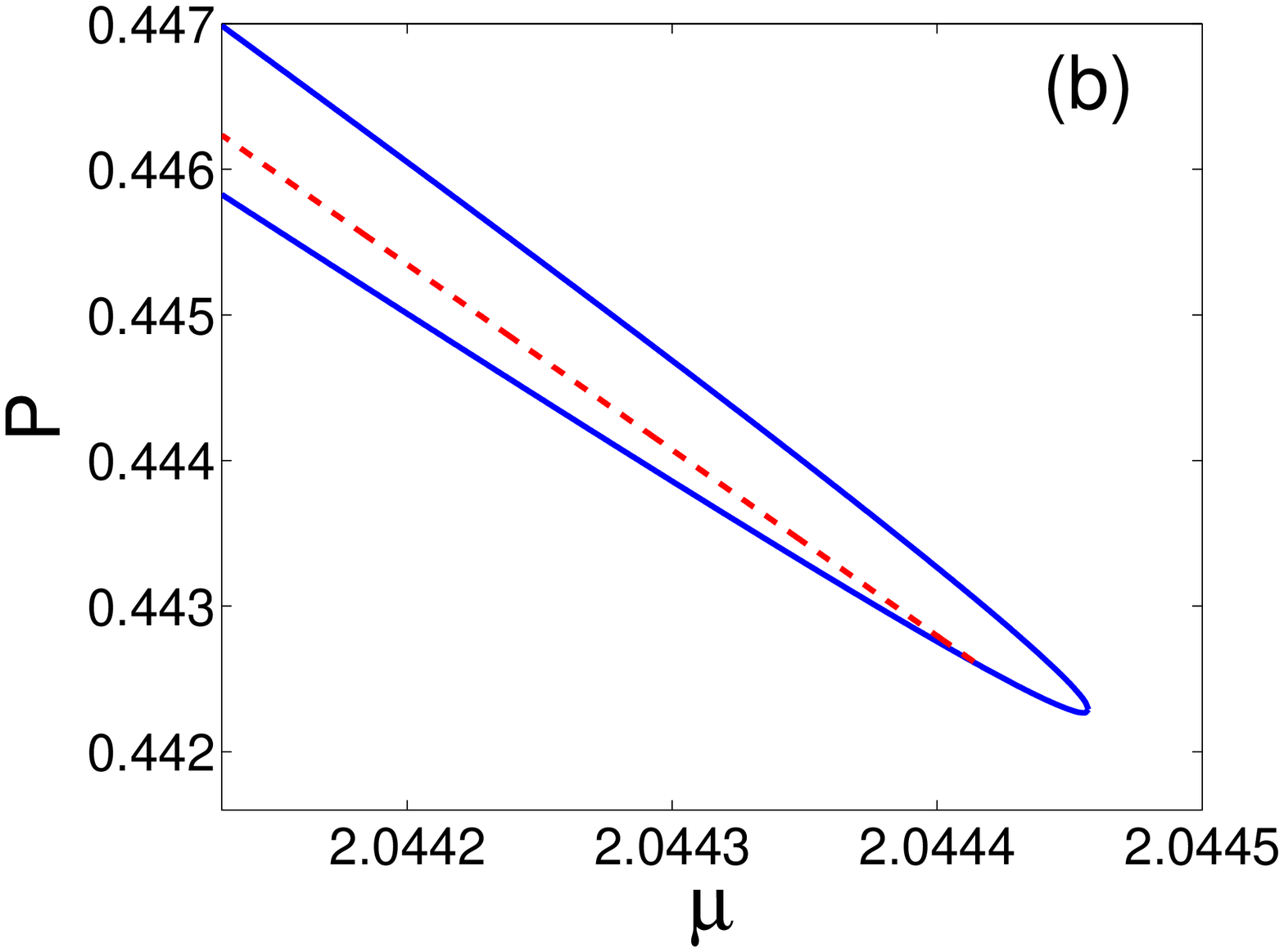} }

\vspace{0.8cm}
\centerline{\includegraphics[width=0.9\textwidth]{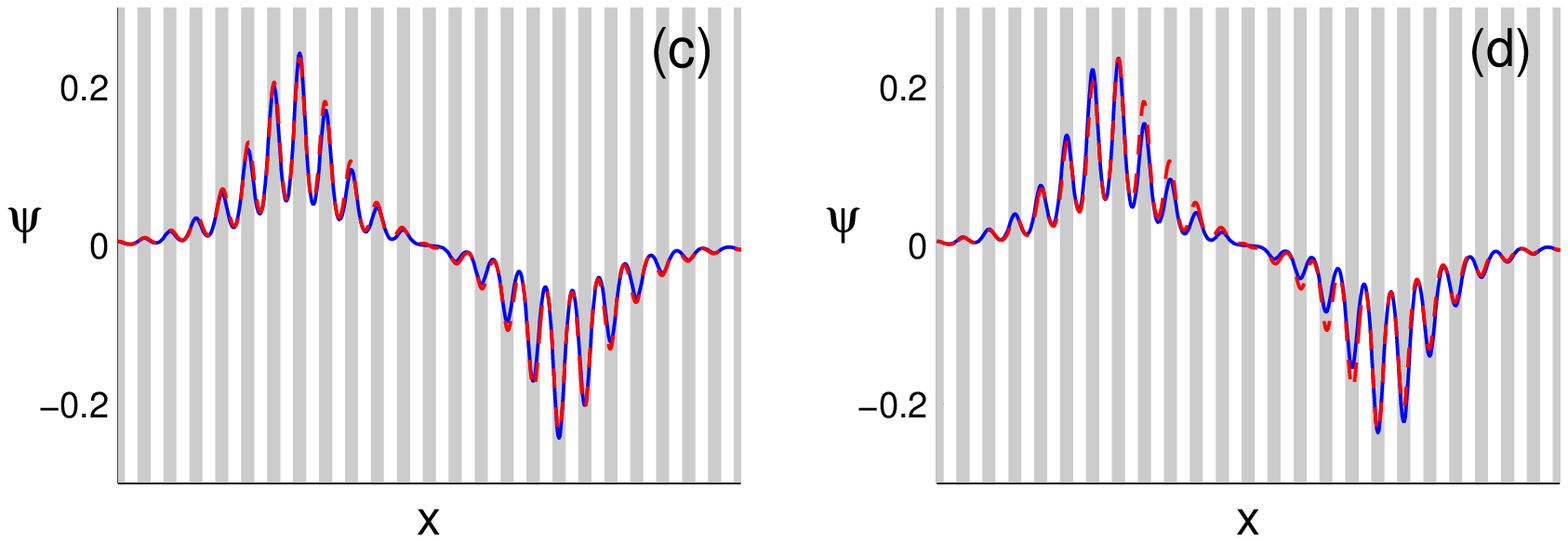} }

\caption{({\it a,b}) Power curves for bound states involving two
solitary waves with opposite envelope polarity and $N=10$, for
self-focusing nonlinearity ($\sigma=1$): ({\it a}) numerical; ({\it
b}) analytical. The analytical solid blue branches are obtained from
equations (\ref{e:powerapproxi2a}) and (\ref{e:conditioncosinea}),
while the analytical dashed red branch is obtained from equations
(\ref{e:powerapproxi2b}) and (\ref{e:conditioncosineb}). ({\it c,d})
Soliton profiles at the points (corresponding to $\mu=2.0438$) of
the lower and upper branches marked by the same letters in ({\it
a}). Solid blue curves: numerical; dashed red curves: analytical as
given by~\eqref{e:psiapproxi}. (Online version in colour.)}
\label{f:opposite_polarity}
\end{figure}

Next, we  turn to quantitative comparison between the analysis and
numerics. For this purpose, we choose to consider bound states
comprising solitary waves with opposite envelope polarity (the
results of comparison for bound states with same envelope polarity
are very similar). Specifically, once again, we assume self-focusing
nonlinearity ($\sigma=1$) and take $N=10$, but the envelopes of the
two solitary waves in the bound state now have opposite polarity.
Figure~\ref{f:opposite_polarity}{\it a} displays the numerical power
diagram near the bifurcation point $\mu_c=2.04442$, and
figure~\ref{f:opposite_polarity}{\it b} illustrates the analytical
power curves given by~\eqref{e:powerapproxi2a}
and~\eqref{e:powerapproxi2b} near the analytical bifurcation point
$\mu_c=\mu_0-\epsilon_c^{(1)2}=2.04446$, where
$\epsilon_c^{(1)}=0.1368$ as obtained from equation
(\ref{e:eqepsilonc}). (To compute the analytical power curves, we first
solve equations~\eqref{e:conditioncosine} for $\delta$ as a function
of $\epsilon$ for $\epsilon_c^{(1)}<\epsilon \ll 1$,  and then use
$\epsilon=\sqrt{|\mu_0-\mu|}$.) Both the numerical and analytical
power curves in figure~\ref{f:opposite_polarity}{\it a,b} exhibit
three branches, as expected. In addition, the middle (mixed-site)
branch now bifurcates from the lower-power (on-site) branch, again in
agreement with the analysis in the end of \S5. Quantitatively, the
numerical and analytical power curves are also in reasonable
agreement, especially considering that the analytical bifurcation
point here is $\epsilon_c^{(1)}=0.1368$, which is not all that
small; moreover, for this value of $\epsilon$ and $N=10$, $\epsilon
S\sim 4$, which is not all that large.  Figure
\ref{f:opposite_polarity}{\it c,d} shows quantitative comparison of
numerics and analysis for the bound-state profiles on
the lower- and higher-power branches at $\mu=2.0438$ (near the
bifurcation point). The corresponding analytical bound-state
profiles (red dashed curves) are given by~\eqref{e:psiapproxi}. Here, the locations of the two solitary waves
involved in the bound state, namely, $x_0^\pm$, are determined
by~\eqref{e:cond2}; these values are dependent on
$\delta(\epsilon)$, which is found by solving
equations~\eqref{e:conditioncosine} with
$\epsilon=\sqrt{|\mu_0-\mu|}$. The analytical solution profiles show
good agreement with the numerical ones.

\begin{figure}[t!]
\centerline{\includegraphics[width=0.9\textwidth]{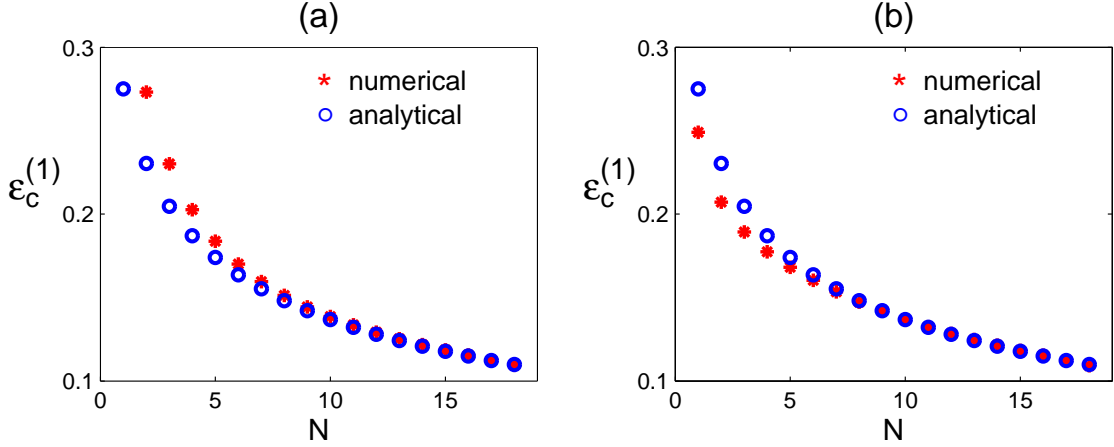}}

\caption{Comparison between analytical and numerical values of the
bifurcation point $\epsilon_c^{(1)}$ at various values of $N$ for
$\sigma=1$ (self-focusing nonlinearity):  ({\it a}) same envelope
polarity; ({\it b}) opposite envelope polarity. The analytical
$\epsilon_c^{(1)}$ is obtained by solving equation
(\ref{e:eqepsilonc}). }   \label{f:epsilonc_compare}
\end{figure}

As another quantitative comparison, we examine the dependence of the
bifurcation point $\epsilon_c^{(1)}$ on the parameter $N$ that
controls the separation distance of the two solitary waves in a
bound state. The analytical and numerical values of this bifurcation
point against $N$ are displayed in figure \ref{f:epsilonc_compare}
for solitons with the same as well as opposite envelope polarities
and self-focusing nonlinearity ($\sigma=1$). Here, the analytical
bifurcation point $\epsilon_c^{(1)}$ is obtained from solving
equation (\ref{e:eqepsilonc}), and this value is the same for both
cases of envelope polarity (with the same $N$). For both envelope
polarities, the numerical values for $\epsilon_c^{(1)}$ approach the
analytical ones when $N\to \infty$, confirming the asymptotic
accuracy of our analysis.

\begin{figure}[t!]
\centerline{\includegraphics[width=0.9\textwidth]{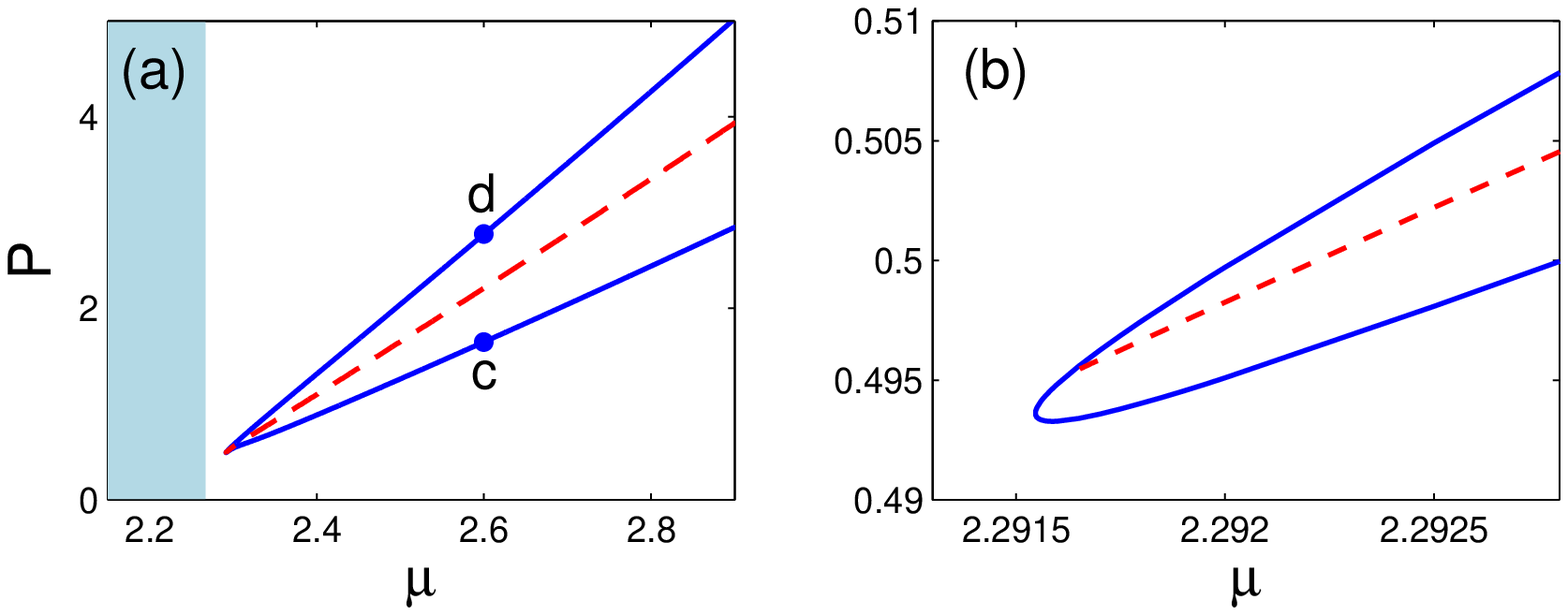}}

\vspace{0.8cm}
\centerline{\includegraphics[width=0.9\textwidth]{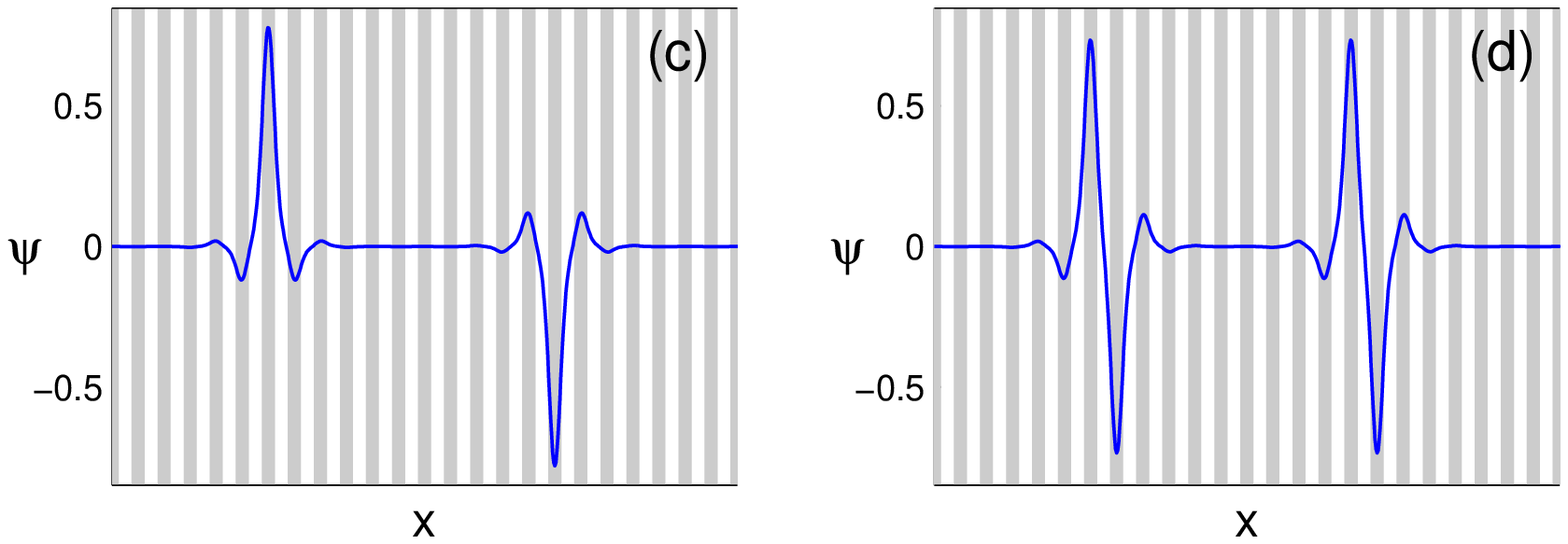}}

\caption{({\it a}) Numerical power curve for bound states involving
solitary waves of the same envelope polarity with $N=10$ and
$\sigma=-1$ (self-defocusing nonlinearity); the shaded region is the
first Bloch band. ({\it b}) Amplification of ({\it a}) near the
bifurcation point. ({\it c,d}) Bound-state profiles at points of the
power curve marked by the same letters in ({\it a}). }
\label{f:same_polarity_def}
\end{figure}

As a final comparison, we assume self-defocusing nonlinearity
($\sigma=-1$) and consider bound states involving solitary waves
with the same envelope polarity. For $N=10$, the numerical results
for this family of bound states are displayed in figure
\ref{f:same_polarity_def}. The power curve features three branches
(figure \ref{f:same_polarity_def}{\it a}), and the middle branch
bifurcates from the upper branch (figure
\ref{f:same_polarity_def}{\it b}), in agreement with the asymptotic
analysis. In addition, the bound state at point c on the lower
branch comprises two on-site solitons which are separated by
$(N+1)\pi=11\pi$ (see figure \ref{f:same_polarity_def}{\it c}), and
the bound state at point d on the upper branch comprises two
off-site solitons which are separated by $N\pi=10\pi$, again
consistent with the previous analysis. It may seem puzzling at first
sight that the two on-site solitons in figure
\ref{f:same_polarity_def}{\it c} are of opposite signs, given that
the envelopes of these solitons ought to have the same polarity. To
explain this feature, note that at the band edge relevant here,
$\mu_0=2.266735$, which is the upper edge of the first Bloch band,
the Bloch mode $p(x)$ is $2\pi$-periodic, and its adjacent peaks
have opposite signs, similar to the function $\cos x$ (see Yang
2010, $\S 6.1.1$). Since the two on-site gap solitons in figure
\ref{f:same_polarity_def}{\it c} are separated by $(N+1)\pi=11\pi$,
the Bloch function $p(x)$ at the centres of these solitons then has
opposite sign. Thus, even though the envelopes of the two gap
solitons have the same polarity, the overall gap-soliton profiles,
which are products of the Bloch function and the envelope function,
have opposite sign, as in figure \ref{f:same_polarity_def}{\it c}.

\section{Concluding remarks}
\label{s:conclusion}

In this paper, an asymptotic theory was developed for stationary
gap-soliton bound states consisting of two fundamental gap solitons
in 1D periodic media. It was shown that there is a countable set of
such bound state families, characterized by the separation distance
of the two solitary waves involved, and each family features three
distinct solution branches that bifurcate near band edges at small,
but finite, amplitude. Of these three solution branches, one branch
contains bound states whose fundamental solitons are both on-site;
the second branch contains bound states whose fundamental solitons
are both off-site; and the third branch represents bound states in
which one fundamental soliton is on-site and the other off-site. The
power curves associated with these solution branches were computed
asymptotically for large solitary-wave separation. On the power
diagram, the on-site solution branch is always the lower-power
branch, the mixed-site branch the middle branch, and the off-site
branch is the higher-power branch; in addition, the middle branch
bifurcates from the upper (lower) branch when the envelopes of the
two fundamental solitons in the bound state have the same (opposite)
polarity. These analytical results were compared with numerical
results and good agreement was obtained.

There are several common features between the results of this
article and those obtained by Yang \& Akylas (1997) for
two-wavepacket bound states in the fifth-order KdV equation. In
that case, a countable set of bound states also bifurcate at finite
values of the wave amplitude, and each solution family also features
multiple branches, with asymmetric-wave branches bifurcating off
symmetric-wave branches. Furthermore, the techniques for
constructing bound states in these two models follow along the same
lines. Even though the current lattice model
(\ref{e:NLSamp}) is not translation invariant, while the fifth-order
KdV equation is, remarkably, both models admit very
similar classes of solitary-wave solutions.

In addition to being of fundamental interest, the bound states
discussed here could also prove useful in applications as they allow
increased flexibility in the profiles of localized nonlinear modes
within band gaps. In a recent demonstration of image transmission in
photonic lattices, in fact, Yang {\it et al.} (2011) utilized 2D
higher-order gap solitons of various shapes, and those gap solitons
are closely related to the 1D bound states constructed in the
present paper. For the purpose of assessing the potential usefulness
of these 1D bound states, of course, it is necessary to examine the
stability properties of the various bound-state families, a question
that will be taken up in future studies.

In this paper, attention was focused on bound states consisting of
only two fundamental gap solitons; extending the analysis to more
complicated bound states involving three or more fundamental
solitons, as well as to bound states in 2D periodic media, will be
left to future studies.

Our final remark is that, since the bound states constructed in this
paper bifurcate near Bloch-band edges at finite amplitude, their
power curves always terminate before band edges and never reach them
(see figures \ref{f:same_polarity}{\it a} and
\ref{f:same_polarity_def}{\it a}). This feature of the power curve
is shared by some other types of gap solitons, such as the
truncated-Bloch-wave solitons (Alexander {\it et al.} 2006; Yang
2010). Whether the present analysis can be applied to those gap
solitons needs further investigation.

\section*{Acknowledgment}

The work of T.R.A. was supported in part by the National Science
Foundation (Grant DMS-098122), and the work of G.H. and J.Y. was
supported in part by the Air Force Office of Scientific Research
(Grant USAF 9550-09-1-0228) and the National Science Foundation
(Grant DMS-0908167).


\catcode`\@ 11
\def\journal#1&#2,#3 {\begingroup \let\journal=\d@mmyjournal {\frenchspacing\it #1\/\unskip\,}
{\bf\ignorespaces #2}\rm, #3\endgroup}
\def\d@mmyjournal{\errmessage{Reference foul up: nested \journal macros}}

\end{document}